\documentclass[acmtog,screen,nonacm]{acmart}
\AtBeginDocument{%
  }

\usepackage{mathtools}
\usepackage{algorithmic}
\usepackage{graphicx}
\usepackage{accents}
\usepackage{textcomp}
\usepackage{algorithmic}
\usepackage[ruled,norelsize]{algorithm2e} 
\usepackage{siunitx}
\usepackage{multirow}
\usepackage{epstopdf}
\usepackage{listings}
\usepackage{glossaries}

\usepackage{listings}
\usepackage{xcolor}

\usepackage{hyperref}
\usepackage{orcidlink}
 
\definecolor{codegreen}{rgb}{0,0.6,0}
\definecolor{codegray}{rgb}{0.5,0.5,0.5}
\definecolor{codepurple}{rgb}{0.58,0,0.82}
\definecolor{codeblue}{rgb}{0.0,0.4,0.8}
\definecolor{backcolour}{rgb}{0.95,0.95,0.92}
 
\lstdefinestyle{mystyle}{
    backgroundcolor=\color{backcolour},   
    commentstyle=\color{codegreen},
    keywordstyle=\bfseries\color{codeblue},%
    numberstyle=\tiny\color{codegray},
    stringstyle=\color{codepurple},
    basicstyle=\ttfamily\footnotesize,
    breakatwhitespace=false,         
    breaklines=true,                 
    captionpos=t,                    
    keepspaces=true,                 
    numbers=none,%
    xleftmargin=1.5em,
    framexleftmargin=1.5em,
    numbersep=5pt,                  
    showspaces=false,                
    showstringspaces=false,
    showtabs=false,                  
    tabsize=2
}

\usepackage{booktabs}
\usepackage[caption=false,font=footnotesize]{subfig}

\usepackage{bm}

\usepackage{pgfplots}
\usepackage{tikz-3dplot}
\pgfplotsset{compat=1.16}
\usetikzlibrary{patterns,external}
\usepgfplotslibrary{colormaps}

\usepackage{tabularx}
\usepackage{amsthm}
\theoremstyle{definition}

\DeclareMathOperator*{\Tx}{Tx}

\newcommand*\diff{\mathop{}\!\mathrm{d}}

\DeclareMathOperator{\Prox}{Prox}

\DeclareMathOperator{\Diag}{Diag}

\DeclareMathOperator*{\argmin}{arg\,min}

\newacronym{pw}{PW}{plane wave}
\newacronym{sg}{SG}{spatial gradient}
\newacronym{tv}{TV}{total vartiation}
\newacronym{sos}{SoS}{speed-of-sound}
\newacronym{masld}{MASLD}{metabolic dysfunction-associated steatotic liver disease}
\newacronym{das}{DAS}{delay-and-sum}
\newacronym{crf}{CRF}{complex radio-frequency}
\newacronym{zlcc}{ZLCC}{zero-lag cross-correlation}
\newacronym{admm}{ADMM}{alternating direction method of multipliers}

\author{Samuel Beuret\textsuperscript{$*$}, Louise Zhuang\textsuperscript{$*$}, Jeremy J. Dahl\textsuperscript{$*$} \MakeLowercase{and} Jean-Philippe Thiran\textsuperscript{$\dagger$}\\
\NoCaseChange{\normalfont\textsuperscript{$*$}Stanford University, Stanford, CA, USA \quad \normalfont\textsuperscript{$\dagger$}EPFL, Lausanne, Switzerland}}

\copyrightyear{2026}
\begin{document}

\title{A Total Variation Regularizer with Partially Known Support for Pulse-Echo Speed-of-Sound Imaging}

\thanks{Samuel Beuret\orcidlink{0000-0001-6558-0892} and Jeremy Dahl\orcidlink{0000-0001-9877-452X} are with the Department of Radiology, Stanford University School of Medicine, Stanford, CA 94305, USA (e-mail: sbeuret@stanford.edu, jjdahl@stanford.edu). Louise Zhuang\orcidlink{0000-0002-5677-4895} is with the Department of Electrical Engineering, Stanford University, Stanford, CA 94305, USA. Jean-Philippe Thiran\orcidlink{0000-0003-2938-9657} is with the Signal Processing Laboratory (LTS5), École Polytechnique Fédérale de Lausanne (EPFL), Lausanne, Switzerland (e-mail: jean-philippe.thiran@epfl.ch) and with the Department of Radiology, University Hospital Center and University of Lausanne (CHUV), Lausanne, Switzerland.}

\begin{abstract}
\noindent\rule{\linewidth}{0.4pt}
    Pulse-echo speed-of-sound imaging is emerging as a powerful extension of conventional echography. It shows particular promise for diagnosing metabolic dysfunction-associated steatotic liver disease, whose prevalence is increasing worldwide alongside obesity rates. However, current techniques often lack robustness, particularly when limited data are available. Existing methods rely on different regularization strategies to incorporate knowledge on the speed-of-sound distribution within tissues. In this work, we propose a novel regularization approach that automatically integrates information from the B-mode image. The proposed regularizer relies on the principle of sparsity with partially known support applied to a total variation norm, enabling increased robustness compared to regularizers disregarding information contained in B-mode images, or circumventing drawbacks of approaches based on the manual segmentation of the latter. We evaluate the proposed technique with simulated and in vivo data, considering both large and reduced datasets, with a specific focus on liver imaging. The results indicate that our method enhances speed-of-sound accuracy with simulated data when compared with regularizers proposed in the literature, whereas the robustness to data reduction is improved with in vivo data. Notably, our approach can facilitate the deployment of pulse-echo speed-of-sound imaging on ultra-portable transducers, which are typically constrained in data acquisition capacity.
  \noindent\rule{\linewidth}{0.4pt}
\end{abstract}

\authorsaddresses{}
\maketitle
\renewcommand{\shortauthors}{Beuret, Zhuang, Dahl, and Thiran}

\section{Introduction}
\label{sec:intro}

The recent years have witnessed the development of pulse-echo \gls{sos} imaging methods. Their aim is to estimate the distribution of local \gls{sos} within the insonified tissues. On the one hand, information regarding the distribution of local \gls{sos} can be used to correct for aberrations in B-mode images caused by \gls{sos} variations \cite{ali_aberration_2023, ali_distributed_2018, rau_ultrasound_2019, jaeger_full_2015}. On the other hand, local \gls{sos} is currently being investigated as a biomarker for certain diseases, which could enhance the diagnostic capabilities of ultrasound imaging. In particular, local \gls{sos} is a promising indicator of \gls{masld}, due to its correlation with the fat fraction in the liver \cite{stahli2023first, imbault_ultrasonic_2018, telichko2022noninvasive}.

Pulse-echo \gls{sos} estimation methods can be classified according to the strength of their assumptions, starting with techniques positing a constant \gls{sos} within the insonified medium \cite{ophir1986estimation, imbault_ultrasonic_2018, bezek2023analytical, xiao2024real}. If they are adapted to certain applications, the usefulness of global \gls{sos} estimation methods in liver imaging is limited since such techniques run the risk of being biased by superficial skin, fat or muscle layers. Layer-based methods---positing constant \gls{sos} along the lateral dimension---aim at circumventing this issue \cite{ali2021local, heriard2023refraction, de2025layer}. They are however limited by the complexity of the media they can image and cannot estimate \gls{sos} (or are biased) in media that are not structured as series of horizontal layers.

In contrast, fully distributed pulse-echo \gls{sos} imaging methods estimate a two-dimensional map of the \gls{sos} of insonified tissues. The majority of current methods follow the same rationale.
Local differences in aberration phase shifts---or, equivalently, aberration delays---are first estimated at different positions within the insonified medium. The aberration delays are then related to a local \gls{sos} map of the tissues through a tomographic linear measurement model. A \gls{sos} map is ultimately computed through the resolution of an inverse problem \cite{jaeger_computed_2015, stahli_improved_2020, ali_sound_2023, sanabria_speed--sound_2018, beuret_windowed_2023-1, sanabria2022direct, jaeger2022pulse}. Different variations to this fundamental pipeline have been proposed, including learning the linear measurement model or its inversion \cite{yolgunlu2024learned, bezek2024learning}, or relying on the matrix imaging framework \cite{bureau2024reflection}.
Moreover, differentiable beamforming---which relies on a similar tomographic measurement model---provides a promising alternative to these techniques that circumvents the need for aberration phase estimation altogether \cite{simson2025ultrasound}.

Unfortunately, fully distributed pulse-echo \gls{sos} imaging methods are inherently hindered in practice by the limited aperture of the ultrasound transducer. Mathematically, this limitation translates into the ill-conditioning of the measurement models detailed above. This issue is akin to the \textit{missing cone problem} encountered in other imaging modalities such as optical tomography \cite{lim2015comparative} or X-ray laminography \cite{holler2019three}, among others. The ill-conditioning of the measurement model introduces a general lack of robustness in the estimation of local \gls{sos}, which is especially prejudicial to imaging with a limited amount of data. In particular, this lack of robustness represents an hindrance to the deployment of pulse-echo \gls{sos} imaging methods onto ultra-portable ultrasound systems.

To circumvent the ill-conditioning of the measurement model and obtain a robust estimation of local \gls{sos}, a strong prior on the \gls{sos} map is therefore necessary. Regularizers proposed in the literature include a Tikhonov regularizer \cite{jaeger_computed_2015, sanabria2022direct}, a regularizer penalizing the squared gradient of the \gls{sos} map (the \gls{sg} regularizer) \cite{stahli_improved_2020}, or a \gls{tv} norm with anisotropic weighting \cite{sanabria_speed--sound_2018}. Learned regularizers are also starting to be investigated \cite{jaeger2023learnt}, along with the regularization prospects of neural implicit representations \cite{byra2024implicit}.

The regularization approaches described up to now do not factor in contextual information regarding the medium in which they are applied. However, the knowledge of the body region being screened or the structures observed in B-mode images contain meaningful information regarding the tissue \gls{sos} distribution we seek to estimate.
An approach to circumvent the lack of robustness in the model inversion is, thus, to employ a conditional \gls{sos} prior dependent on the content of the B-mode image. An example of such a method is proposed by Stähli \textit{et. al.} \cite{stahli_bayesian_2021} in the context of liver \gls{sos} imaging. The insonified medium is first segmented by a practitioner on the basis of a B-mode image of the tissues. A Bayesian soft prior is then introduced to ensure that the \gls{sos} is approximately constant in the different segmented zones of the medium. Unfortunately, the need for manual segmentation of B-mode images by a trained practitioner significantly hinders the applicability of the technique.

Our aim in this article is to propose a different mechanism to automatically factor in information from a B-mode image into a local \gls{sos} regularizer. In particular, we focus on liver imaging due to the potential use of \gls{sos} as a biomarker for \gls{masld}. In ultrasound liver imaging, the boundaries between tissues---skin, muscle, fat or hepatic tissue---generally generate bright ridges in B-mode images. This is especially true for the liver capsule, which often provides a clear delimitation to the organ. The proposed method is thus based on the following rationale. First, we detect ridges in a B-mode image of the tissues. Second, we introduce a regularizer on the local \gls{sos} favoring piecewise-constant maps whose edges coincide with the ridges of the B-mode image. Finally, we determine local \gls{sos} through an inverse problem taking into account the regularizer and the aberration phase shifts estimated according to the method outlined in \cite{beuret_windowed_2023-1}.

The article is organized as follows. First, we describe the different steps of the proposed method, including data acquisition, aberration delay estimation, ridge detection, the inverse problem used for \gls{sos} estimation and the associated optimization algorithm. Second, we describe the experiments performed to test the effectiveness of the proposed approach, as well as their results with simulated and in vivo data. We follow with a discussion of the results before concluding.

\section{Method}

\subsection{Aberration Phase Shifts}

The first step of the proposed method is to estimate local aberration phase differences at different points of the insonified medium. There are several approaches proposed in the literature to do so (e.g., \cite{stahli_improved_2020, ali_aberration_2023}, or \cite{sanabria2022speed}). However, we focus in this work on the method proposed in \cite{beuret_windowed_2023-1} because of its empirical performance with small numbers of insonifications. We recall in this section the main steps of the latter aberration phase shift estimation method and refer the reader to \cite{beuret_windowed_2023-1} for additional details regarding the implementation and parameters of the technique.

\subsubsection{Data acquisition}

The first step of the aberration phase shift estimation process is to emit a series of $N^{\text{Tx}}$ steered \glspl{pw} with a linear transducer and to acquire for each insonification the \gls{crf} signals associated with every transducer element.

\subsubsection{Beamforming}

The second step consists in reconstructing $N^{\text{Tx}}$ \gls{crf} images---one per insonification---using \gls{das}, assuming a constant tissue \gls{sos} $c_0$. Then, one must construct a series of $N^{\text{dif}}$ \gls{crf} images characterized by a constant difference between the emission angle (the steering angle of the \gls{pw}) and the angle at which the scattered waves reach the transducer \cite{heriard2023refraction}, the set of $N^{\text{dif}}$ images being parametrized by $\boldsymbol{\theta}^{\text{dif}} \in \mathbb{R}^{N^{\text{dif}}}$ a vector of the \textit{dif angles}, namely the halved angle differences. In practice, this is achieved by filtering in the spatial Fourier domain the $N^{\Tx}$ \gls{crf} single-\gls{pw} images \cite{stahli_improved_2020}, followed by a weighted sum of the filtered images.

\subsubsection{Windowed Radon transform}

Once we obtain a series of $N^{\text{dif}}$ images with constant differences between emission and reception angles, we propose to perform the windowed Radon transform of the image. Practically, we window the $N^{\text{dif}}$ images around a grid of points of size $N^{\text{x,} \Delta \phi} \times N^{\text{z,} \Delta \phi}$. We then compute the Radon transform of the $N^{\text{dif}} N^{\text{x,} \Delta \phi} N^{\text{z,} \Delta \phi}$ resulting patches, considering a vector $\boldsymbol{\theta}^{\text{mid}} \in \mathbb{R}^{N^{\text{mid}}}$ of angles used in the Radon transform. The result is, for a single patch, a series of $N^{\text{mid}}$ single-dimensional signals indexed by their Radon angle.

\subsubsection{Aberration phase shift estimation}

The aberration phase shifts are then estimated by computing the phase of the \gls{zlcc} of signals sharing the same position and Radon angle, but differing by which of the $N^{\text{dif}}$ constant dif-angle images they are taken from. In practice \gls{zlcc} is performed between signals corresponding to adjacent dif-angles, and a subset of $N^{\phi}$ phase shift maps is selected among all $N^{\text{mid}}\big( N^{\text{dif}} - 1\big)$ possibilities. All the resulting aberration phase shift map subsets are ultimately gathered and vectorized as $\boldsymbol{\Delta \phi} \in \mathbb{R}^{N^{\text{d}}}$, with $N^{\text{d}} = N^{\phi} N^{\text{x,} \Delta \phi} N^{\text{z,} \Delta \phi}$. We can associate to $\boldsymbol{\Delta \phi}$ a vector $\boldsymbol{w} \in \{ 0,1\}^{N^{\text{d}}}$ indicating if the phase shift estimate is valid or not, according to if the estimation is performed within the support of the emitted and received \glspl{pw}.

\subsection{Ridge Detection}
\label{sec:meth_ridge}

The proposed approach requires detecting interfaces between tissues, identified as ridges in a B-mode image. Ridge detection is a well-mastered technique in image analysis \cite{eberly_ridges_1994, jacob_design_2004}.
Its applications to medical imaging include the detection of vessels in color images and angiographic MRI \cite{frangi_multiscale_1998, frangi_model-based_1999, staal_ridge-based_2004}, or the detection of neurites in fluorescence microscopy \cite{meijering_novel_2003}, among others. Following~\cite{meijering_novel_2003,jacob_design_2004}, we rely herein a simple approach to detect ridges in an ultrasound B-mode image.

First, we reconstruct a B-mode image of the medium with traditional \gls{das} and coherent compounding \cite{montaldo_coherent_2009}, using the data obtained from the $N^{\text{Tx}}$ \gls{pw} insonifications. We posit that a B-mode image of the medium---obtained after envelope detection and log-compression---is represented by a continuous function $y^{\text{B}}(\boldsymbol{r})$. The first step of the ridge detection method is to compute its Hessian matrix at each point $\boldsymbol{r}$ of the medium. To guarantee a sufficient robustness, we rather compute the Hessian of a smoothed version of $y^{\text{B}}$ obtained by convolving the image with a Gaussian filtering kernel.
Therefore, we express the Hessian of $\gamma^{\text{B}}$ as
\begin{equation}
    \renewcommand*{\arraystretch}{1.2}
	\boldsymbol{H}^{\text{B}} \big(\boldsymbol{r}\big) =
	\begin{bmatrix}
		\big[ h^{\text{xx}} *_{\boldsymbol{r}} y^{\text{B}} \big]\big(\boldsymbol{r}\big) &
		\big[ h^{\text{xz}} *_{\boldsymbol{r}} y^{\text{B}} \big]\big(\boldsymbol{r}\big)\\
		\big[ h^{\text{xz}} *_{\boldsymbol{r}} y^{\text{B}} \big]\big(\boldsymbol{r}\big) &
		\big[ h^{\text{zz}} *_{\boldsymbol{r}} y^{\text{B}} \big]\big(\boldsymbol{r}\big) 
	\end{bmatrix},
\end{equation}
\begin{equation}
\begin{gathered}
	h^{\text{xx}}\big(\boldsymbol{r}\big) = \frac{\diff^2}{\diff{x}^2} g_{\sigma} \big(\boldsymbol{r} \big),\quad h^{\text{zz}} \big(\boldsymbol{r} \big) = \frac{\diff^2}{\diff{z}^2} g_{\sigma} \big(\boldsymbol{r} \big),\\
	h^{\text{xz}} \big(\boldsymbol{r} \big) = \frac{\diff^2}{\diff{x}\diff{z}} g_{\sigma} \big(\boldsymbol{r} \big),
 \end{gathered}
\end{equation}
where $*_{\boldsymbol{r}}$ denotes a spatial convolution and $g_{\sigma}$ is an isotropic Gaussian distribution with standard deviation $\sigma$. The minimum eigenvalue of $\boldsymbol{H}^{\text{B}}$ at each point of the medium is denoted by $\lambda^{\text{min}}\big(\boldsymbol{r} \big)$. A negative value of $\lambda^{\text{min}}\big(\boldsymbol{r} \big)$ indicates the possible presence of a ridge, and the probability that there is a ridge in $\boldsymbol{r}$ improves as $\lambda^{\text{min}}\big(\boldsymbol{r} \big)$ lowers. Therefore, we apply thresholding to $-\lambda^{\text{min}}(\boldsymbol{r})$ to obtain a mask $q$ indicating the presence of ridges in the image. Mathematically, we write
\begin{equation}
	q \big(\boldsymbol{r} \big) = H\Big( -\lambda^{\text{min}}\big(\boldsymbol{r} \big) - \lambda^{\text{thresh}}\Big),
\end{equation}
with $H$ the Heaviside step function. Lastly, we must highlight that the approach used herein, where estimation of the Hessian of the image and eigenvalue processing are required, shares strong conceptual similarities with anisotropic diffusion techniques for speckle reduction \cite{finn2011echocardiographic, abd2002real, krissian2007oriented}.

In our implementation of ridge detection, the B-mode image is represented by a vector $\boldsymbol{y}^{\text{B}} \in \mathbb{R}^{N^{\text{x, B}} N^{\text{z, B}}}$, with $N^{\text{x, B}}$ and $N^{\text{z, B}}$ the numbers of pixels along the $\text{x}$ and $\text{z}$ directions. Filtering by $h^{\text{x}\text{x}}$, $h^{\text{x}\text{z}}$, and $h^{\text{z}\text{z}}$ is performed using discrete convolutions. Ultimately, we compute a matrix $\boldsymbol{H}^{\text{B}}$ for each pixel in the image, and functions $\lambda^{\text{min}}$ and $q$ are represented by vectors $\boldsymbol{\lambda}^{\text{min}} \in \mathbb{R}^{N^{\text{x, y}} N^{\text{z, y}}}$ and $\boldsymbol{q} \in \{0,1\}^{N^{\text{x, y}} N^{\text{z, y}}}$, respectively. 

\subsection{Inverse Problem}
\label{sec:meth_inv}

We need to estimate the \gls{sos} map from the aberration phase shift maps $\boldsymbol{\Delta \phi}$, taking into account information from the ridge mask $\boldsymbol{q}$. To do so, we follow \cite{jaeger_computed_2015, stahli_improved_2020, beuret_windowed_2023-1} and introduce first the slowness difference map $\boldsymbol{\Delta s} \in \mathbb{R}^{\text{s}}$ defined as
\begin{equation}
\label{eq:c_ds}
    \boldsymbol{\Delta s} = \mathbf{1} \oslash \boldsymbol{c} - \frac{1}{c_0},
\end{equation}
with $c_0$ the \gls{sos} assumed during beamforming and $\oslash$ the element-wise division between two vectors. In \eqref{eq:c_ds}, $\boldsymbol{c} \in \mathbb{R}^{N^{\text{s}}}$ denotes a vectorization of the local \gls{sos} map, with $N^{\text{s}}=N^{\text{x}}N^{\text{z}}$, where $N^{\text{x}}$, $N^{\text{z}}$ are the numbers of pixels along the $\text{x}$ and $\text{z}$ axes, respectively.
As detailed in \cite{beuret_windowed_2023-1}, the vector of phase-shift maps $\boldsymbol{\Delta \phi}$ can be estimated---within a set of valid elements indicated by $\boldsymbol{w} \in \{0,1\}^{N^{\text{d}}}$---as the product of $\boldsymbol{\Delta s}$ by a discrete measurement model $\boldsymbol{M} \in \mathbb{R}^{N^{\text{d}} \times N^{\text{s}}}$. 

The measurement model $\boldsymbol{M}$ is defined as the concatenation of the discretized version of the continuous operators
\begin{multline}
    \label{eq:meth_mod_con}
    \mathcal{M}_k = \frac{\omega_0}{\cos \big( \theta^{\text{dif, +}}_k\big)} \left[\mathcal{T}_{\theta^{\text{mid}}_k + \theta^\text{dif, +}_k} + \mathcal{T}_{\theta^{\text{mid}}_k - \theta^\text{dif, +}_k}\right] -\\ \frac{\omega_0}{\cos \big( \theta^{\text{dif, -}}_k\big)} \left[\mathcal{T}_{\theta^{\text{mid}}_k + \theta^\text{dif, -}_k} + \mathcal{T}_{\theta^{\text{mid}}_k - \theta^\text{dif, -}_k}\right],
\end{multline}
for each $k = 1, ..., N^{\phi}$, where $\omega_0$ is the center frequency of the probe expressed in \SI{}{\radian \per \second}. In \eqref{eq:meth_mod_con}, $\theta^{\text{mid}}_k$, $\theta^{\text{dif, +}}_k$, and $\theta^{\text{dif, -}}_k$ denote the Radon angle associated with $k$, as well as the dif angles parameterizing the images compared to obtain the $k$-th phase shift map, respectively. Operator $\mathcal{T}$ acts on a continuous slowness difference map $\Delta s$ and is defined for an arbitrary angle $\theta$ as
\begin{equation}
    \mathcal{T}_{\theta}\{\Delta s\}(x, z) = \int_{-z}^{0} \Delta s \big(x + \tan(\theta)\hat{z}, \hat{z}+z \big) \frac{1}{\cos(\theta)}\diff{\hat{z}}.
\end{equation}

To factor in information from the ridge map, we take inspiration from the literature on compressed sensing with partially known support \cite{vaswani_modified-cs_2010, jacques_short_2010, carrillo_iterative_2010}. In particular, we penalize the gradient of the slowness difference map only in points $\boldsymbol{r}$ where $q \big( \boldsymbol{r} \big) = 0$.
Mathematically, we thus update the optimization problem compared to the one proposed in \cite{beuret_windowed_2023-1} as
\begin{multline}
\label{eq:ch_SoS_sec_TV_gen}
    \min_{\boldsymbol{\Delta s} \in \mathbb{R}^{N^{\text{s}}}} 
    \frac{1}{2} \big\| \Diag(\boldsymbol{w}) \boldsymbol{M} \boldsymbol{\Delta s} - \Diag(\boldsymbol{w}) \boldsymbol{\Delta \phi} \big\|^{2} +\\
    \sum_{k=1}^{N^{\text{r}}} 
    \big(1 - q^{\Delta s}_k \big) \sqrt{ \lambda^{\text{x}} \left[ \boldsymbol{D}^{\text{x}} \boldsymbol{\Delta s} \right]^2_{k} + \lambda^{\text{z}} \left[  \boldsymbol{D}^{\text{z}} \boldsymbol{\Delta s} \right]^2_{k}}.
\end{multline}
In \eqref{eq:ch_SoS_sec_TV_gen}, $\boldsymbol{D}^{\text{x}} \in \mathbb{R}^{N^{\text{r}} \times N^{\text{s}}}$ and $\boldsymbol{D}^{\text{z}}  \in \mathbb{R}^{N^{\text{r}} \times N^{\text{s}}}$ denote the forward finite difference operators along the $\text{x}$ and $\text{z}$ axes, with $N^{\text{r}}=(N^{\text{x}} -1)(N^{\text{z}} -1)$.
Regularization parameters $\lambda^{\text{x}}$ and $\lambda^{\text{z}}$ quantify the strength of the regularizer along both the $x$ and $z$ axes. Vector $\boldsymbol{q}^{\Delta s} \in \mathbb{R}^{N^{\text{r}}}$ is a version of $\boldsymbol{q}$ interpolated to fit the output dimensions of $\boldsymbol{D}^{\text{x}}$ and $\boldsymbol{D}^{\text{z}}$. Therefore, the regularizer introduced in \eqref{eq:ch_SoS_sec_TV_gen} favors piece-wise constant slowness maps whose edges concur with the ridges of the B-mode image.

State-of-the-art regularizers proposed for pulse-echo \gls{sos} imaging---including the \gls{sg} and \gls{tv} norm approaches---are designed to enforce a uniform \gls{sos} within each distinct tissue of the medium. Unfortunately, they also have a tendency to bias the \gls{sos} estimated in a given tissue towards the \gls{sos} of the adjacent ones. The regularization parameters must therefore achieve a critical trade-off between uniformity and bias. In contrast, the key advantage of the proposed regularizer lies in its strong mitigation of cross-tissue bias. Specifically, it avoids penalizing the gradient of the slowness difference map $\boldsymbol{\Delta s}$ at non-zero pixels of the edge support map $\boldsymbol{q}^{\Delta s}$, which identify the transitions between different tissues. Moreover, the regularization parameters $\lambda^{\text{x}}$ and $\lambda^{\text{z}}$ can be significantly increased compared to the \gls{sg} or \gls{tv} approaches, promoting greater tissue-wise uniformity and yielding more robust \gls{sos} estimates. In the specific case of liver imaging, where tissues are globally organized as horizontal layers, the reduced risk in cross-tissue bias offered by the proposed regularizer allows us to increase $\lambda^{\text{z}}$ further relative to $\lambda^{\text{x}}$, thereby enforcing stronger tissue homogeneity in the axial direction.

\subsection{Optimization Algorithm}

Problem~\eqref{eq:ch_SoS_sec_TV_gen} does not admit a closed-form solution and we need to rely on an iterative algorithm. In addition, there are strong incentives to limit the convergence time of the algorithm to ease the use of pulse-echo \gls{sos} imaging in a clinical context. To obtain such an algorithm, we first rewrite \eqref{eq:ch_SoS_sec_TV_gen} as
\begin{equation}
\label{eq:ch_SoS_sec_inv_prior_2}
\min_{\boldsymbol{\Delta s} \in \mathbb{R}^{N^{\text{s}}} ,\ \boldsymbol{\zeta} \in \mathbb{R}^{N^{\text{d}} + 2 N^{\text{r}}}} \mathcal{F}(\boldsymbol{\zeta}) \text{ s.t. } \boldsymbol{A}\; \boldsymbol{\Delta s} - \boldsymbol{b} = \boldsymbol{\zeta},
\end{equation}
with
\begin{equation}
\boldsymbol{A} = \Big[ \Diag(\boldsymbol{w}) \boldsymbol{M}, \sqrt{\lambda^\text{x}} \boldsymbol{D}^\text{x}, \sqrt{\lambda^\text{z}} \boldsymbol{D}^\text{z} \Big],
\end{equation}
\begin{equation}
\boldsymbol{b} = 
\begin{bmatrix}
\Diag(\boldsymbol{w}) \boldsymbol{\Delta \phi}\\
\boldsymbol{0}\\
\boldsymbol{0}
\end{bmatrix},
\end{equation}

\begin{equation}
\label{eq:f_def}
    \mathcal{F}(\boldsymbol{\zeta}) =  \frac{1}{{2}} \big\|\boldsymbol{\zeta}_1 \big\|^{2} + \sum_{k=1}^{N^\text{r}} \big[1 - q^{\Delta s}_k \big] \sqrt{\left[ \boldsymbol{\zeta}_2 \right]_k^2 + \left[ \boldsymbol{\zeta}_3 \right]_k^2 },
\end{equation}
and
\begin{equation}
    \boldsymbol{\zeta} =
    \begin{bmatrix}
    \boldsymbol{\zeta}_1\\
    \boldsymbol{\zeta}_2\\
    \boldsymbol{\zeta}_3
    \end{bmatrix},
\end{equation}
an auxiliary variable.

We propose to resolve \eqref{eq:ch_SoS_sec_inv_prior_2} using the \gls{admm}\cite{neal2011distributed}, where the \gls{admm} parameter is set to 1. The resulting optimization scheme is detailed in Algorithm \ref{alg:ch_SoS_sec_TV_ADMM}, where $N^{\text{iter}}$ and $\epsilon$ denote the maximum number of iterations and the stopping criterion, respectively. The proximal operator of $\mathcal{F}$  in Algorithm \ref{alg:ch_SoS_sec_TV_ADMM} ($\Prox_{\mathcal{F}}$) admits a computationally cheap closed-form solution that we describe in detail in the Appendix. 
Moreover, computing $\boldsymbol{\Delta s}_{k+1}$ in the main loop of Algorithm \ref{alg:ch_SoS_sec_TV_ADMM} is equivalent to the inverse problem used in \cite{beuret_windowed_2023-1}. In particular, we can precompute the matrix inverse such that computing $\boldsymbol{\Delta s}_{k+1}$ requires only matrix-vector products, leading to a limited time per iteration in practice.

\begin{algorithm}[h]
    \caption{ADMM for problem \eqref{eq:ch_SoS_sec_TV_gen}}
    \label{alg:ch_SoS_sec_TV_ADMM}
    \vspace{0.25\baselineskip}
    $\displaystyle \boldsymbol{\Delta s}_1 \gets \boldsymbol{0}$;\\
    $\displaystyle \boldsymbol{\zeta}_1 \gets \boldsymbol{0}$;\\
    $\displaystyle \boldsymbol{u}_1 \gets \boldsymbol{0}$;\\
    \For{$\displaystyle k := 1$ \text{to} $\displaystyle N^{\textnormal{iter}}$}
    {
        $\displaystyle \boldsymbol{\Delta s}_{k+1} \gets \big[ \boldsymbol{A}^T \boldsymbol{A} \big]^{-1} \boldsymbol{A}^T \big[ \boldsymbol{b} + \boldsymbol{\zeta}_k - \boldsymbol{u}_k \big]$;\\
        \vspace{0.25\baselineskip}
        $\displaystyle \boldsymbol{\zeta}_{k+1} \gets \Prox_{\mathcal{F}} \big(\boldsymbol{A}\; \boldsymbol{\Delta s}_{k+1} - \textbf{b} +\boldsymbol{u}_k \big)$;\\
        \vspace{0.25\baselineskip}
        $\displaystyle \boldsymbol{u}^{k+1} = \boldsymbol{u}_{k} + \big[ \boldsymbol{A}\; \boldsymbol{\Delta s}^{k+1} - \boldsymbol{\zeta}_{k+1} - \boldsymbol{b} \big]$;\\
        \If{$\displaystyle \big\|\boldsymbol{\Delta  s}_{k+1} - \boldsymbol{\Delta s}_{k} \big\| < \epsilon$}{
            \vspace{0.25\baselineskip}
            \textbf{Break}
        }
    }
    \textbf{Return:}\ $\boldsymbol{\Delta s}_{k+1}$
\end{algorithm}

\section{Experiments}

\subsection{System and Sequence}

To test the proposed method, we consider a GE9L-D linear ultrasound transducer (GE Healthcare, Chicago, IL, USA) connected to a Vantage 256 acquisition system (Verasonics, Kirkland, WA, USA). The transducer is composed of 192 piezoelectric elements with a central frequency  of \SI{5.3}{\mega \hertz} and a pitch of \SI{230}{\micro \meter}. The local aberration phase estimation method we employ to assess the proposed regularizer requires the emission of a series of steered \glspl{pw}. Therefore, we consider the data obtained from 115 \gls{pw} insonifications of the medium with steering angles ranging from -28.5° to 28.5° with a 0.5° angular spacing. In addition, we test the proposed regularizer with a reduced amount of data. Thus, we downsample the full 115-\gls{pw} dataset by a factor 10 to simulate data acquisition using only 11 steered \glspl{pw} ranging from -25° to 25° with a 5° angular spacing. Such large ranges of steering angles are necessary to mitigate the missing cone problem highlighted in Section \ref{sec:intro} and allow for a sufficiently robust local \gls{sos} estimation.

\subsection{Simulated Data}

We first assess the performance of the proposed regularizer using simulations of three numerical \gls{sos} phantoms, denoted S1, S2, and S3. These phantoms are comprised of several superficial layers imitating possible configurations of the abdominal wall, while the bottom of the medium is occupied by a uniform area with a \gls{sos} of \SI{1560}{\meter \per \second}. We simulate for each phantom  the emission of 115 \glspl{pw} using k-Wave \cite{treeby_k-wave_2010}, as well as the acquisition of the echo signals by a geometrical model of the transducer. In all the experiments, we posit that the medium possesses a uniform echogenicity.

In vivo measurements suffer in practice from a significant amount of phase noise. This noise is typically generated by movements of the tissues and the transducer, or small-scale \gls{sos} variations overlooked by the proposed model. To improve the verisimilitude of this first set of simulated data, we generate noisy versions of datasets S1 to S3 by multiplying, for each insonification and piezoelectric element, the simulated \gls{crf} signals with the complex exponential of a phase noise term. We define the phase noise as a smoothed Gaussian noise whose amplitude increases with the propagation time. In the following, we present the results of experiments with both noiseless and noisy data to quantify the effect of phase noise on the accuracy of the results.

Our aim with datasets S1 to S3 is to test the dependence of the proposed regularizer to the edge support map. To do so, we define for each dataset three different edge support maps $\boldsymbol{q}$. The first one---denoted as \textit{support 1}---corresponds to edges of the ground-truth \gls{sos} map where some sections are removed. The second one---denoted as \textit{support 2}---coincides exactly with the edges of the \gls{sos} map. The third one---denoted as \textit{support 3}---supposes that additional edges not corresponding to actual \gls{sos} transitions are present. It is intended to represent the most realistic situation.

In addition, we want to assess the minimum contrast that can be achieved by the proposed regularizer. To do so, we simulate an additional set of data using a \SI{1540}{\meter \per \second} medium with an horizontal layer---denoted as \textit{layer 1}---located between depths of \SI{12}{\milli \meter} and \SI{17}{\milli \meter} whose \gls{sos} $c_1$ vary between \SI{1540}{\meter \per \second} and \SI{1560}{\meter \per \second} with a step size of \SI{5}{\meter \per \second}. For each value of $c_1$, we keep the same specifications than the ones assumed in the previous experiments, including the three types of synthetic edge maps support 1, support 2, and support 3.

In the experiments described above, the positions of the edges in the support maps coincides with the physical positions of the \gls{sos} transitions. However, in a realistic setting, timing errors caused by \gls{sos} variations shift the positions of the detected edges compared to their true positions. Moreover, echogenicity variations in the medium also induce specific artifacts in the phase shift maps, thus affecting \gls{sos} estimation. To test the influence of these factors on \gls{sos} estimation, we simulate two additional numerical phantoms, denoted S4 and S5, where both the ground-truth \gls{sos} and echogenicity maps intend to model the anatomy of the abdominal wall. S4 models a series of fat and muscle layers above the liver, whereas S5 models the presence of a thick layer of superficial fat above other subcutaneous tissues. The latter case is intended to study the applicability of the technique in overweight patients. For experiments with S4 and S5, we estimate the edge support map using the algorithm detailed in Section~\ref{sec:meth_ridge}.

Finally, to obtain a quantitative criterion for the effectiveness of a regularizer with simulated data, we compute the median absolute error between ground-truth \gls{sos} maps and the estimated ones.

\subsection{In Vivo Data}

In addition, we test the proposed method on a series of $6$ in vivo datasets---denoted I1 to I6. We acquired them at different positions on the abdominal wall of a healthy volunteer (male, 27 years old). The volunteer provided informed consent, and the experiments were conducted in accordance with the Declaration of Helsinki. Each dataset features a series of superficial skin, fat and muscle layers on top of the liver. In particular, the liver is situated at relatively low depth with cases I1 to I4, but its depth increases with cases I5 and, to a higher degree, I6. We estimate the edge support maps with the method proposed in Section~\ref{sec:meth_ridge} using a B-mode image of the medium reconstructed from the 11-\gls{pw} datasets. 

The method presented in this article is targeted at the estimation of \gls{sos} in the liver for the diagnosis of \gls{masld}. With this aim in mind, we propose to compute the median liver \gls{sos} obtained with the proposed and reference regularizers. We can then analyze the differences in liver \gls{sos} obtained with 115 \glspl{pw} and 11 \glspl{pw}, respectively, to estimate the robustness of the regularization strategies to reductions in the number of insonifications in a realistic use case.

\subsection{Parameters}
\label{sec:exp_param}

The method proposed herein relies on a series of parameters that must be adapted to the system, sequence and use case.
First, we empirically set the two parameters of the ridge detection method to $\sigma=$ \SI{400}{\micro \meter} and $\lambda^{\text{thresh}}=$ 0.025. The values of the two parameters are selected to enforce a qualitative trade-off between precision and recall when the method is applied to the in vivo images. Second, we define a spatial grid spacing of \SI{600}{\micro \meter} along both the x and z directions for the \gls{sos} map, whereas we set the size of the grid to $N^{\text{x}}=71$ and $N^{\text{z}}=83$, resulting in an imaging depth of just below \SI{5}{\centi \meter}. Third, the parameters of the phase estimation process are otherwise identical to the one used in \cite{beuret_windowed_2023-1}, and we refer the reader to the latter article for additional details.

However, the most critical parameters to determine are the two regularization terms $\lambda^{\text{x}}=5\cdot 10^{12}$ and $\lambda^{\text{z}}=5\cdot 10^{12}/5$. In practice, we estimate their optimal values using \gls{sos} maps reconstructed from in vivo datasets I1 to I6. These values were determined empirically, based on a qualitative trade-off between observed intra-tissue \gls{sos} uniformity and the risk of cross-tissue \gls{sos} correlation that may arise due to inaccuracies in the edge support map. This empirical strategy contrasts with a fully quantitative approach, which would involve measuring discrepancies between estimated and ground-truth \gls{sos} maps (e.g., using in vitro phantom data or simulated data). We chose the former one due to the substantial practical challenges involved in acquiring (or simulating) data with known ground truth that would reflect both a realistic tissue structure and a plausible noise patterns in the data (e.g. movement-induced phase noise or multiple scattering artifacts).

Finally, we set the stopping criterion of Algorithm \ref{alg:ch_SoS_sec_TV_ADMM} to $\epsilon=6\cdot 10^{-4}$ and the maximum number of iterations to $N^{\text{iter}}=150$.

\subsection{Reference methods}

We compare the proposed prior with other regularizers introduced in the pulse-echo \gls{sos} imaging literature. First, we consider the \gls{sg} approach used in \cite{stahli_improved_2020}. This approach penalizes anisotropically the squared $\ell_2$ norm of the gradient of the slowness difference map $\boldsymbol{\Delta s}$. The corresponding inverse problem is thus defined as
\begin{multline}
    \label{eq:ex_sg}
    \min_{\boldsymbol{\Delta s} \in \mathbb{R}^{N^{s}}} 
    \frac{1}{2} \big\| \Diag(\boldsymbol{w}) \boldsymbol{M} \boldsymbol{\Delta s} - \Diag(\boldsymbol{w}) \boldsymbol{\Delta \phi} \big\|^{2} +\\
    \frac{\lambda^{\text{x}}}{2} \left\| \boldsymbol{D}^{\text{x}} \boldsymbol{\Delta s} \right\|^2 + \frac{\lambda^{\text{z}}}{2} \left\|  \boldsymbol{D}^{\text{z}} \boldsymbol{\Delta s} \right\|^2.
\end{multline}
A closed-form solution to \eqref{eq:ex_sg} exists and $\boldsymbol{\Delta s}$ can be computed through direct matrix multiplication.

We also take into account an anisotropically weighted TV norm, similar in nature to the one used in \cite{sanabria_speed--sound_2018}. The corresponding inverse problem is defined as
\begin{multline}
\label{eq:ex_tv}
    \min_{\boldsymbol{\Delta s} \in \mathbb{R}^{N^{s}}} 
    \frac{1}{2} \big\| \Diag(\boldsymbol{w}) \boldsymbol{M} \boldsymbol{\Delta s} - \Diag(\boldsymbol{w}) \boldsymbol{\Delta \phi} \big\|^{2} +\\
    \sum_{k=1}^{N^{\text{r}}} \sqrt{ \lambda^{\text{x}} \left[ \boldsymbol{D}^{\text{x}} \boldsymbol{\Delta s} \right]^2_{k} + \lambda^{\text{z}} \left[  \boldsymbol{D}^{\text{z}} \boldsymbol{\Delta s} \right]^2_{k}}.
\end{multline}
Namely, we recover problem \eqref{eq:ch_SoS_sec_TV_gen} where $q^{\Delta s}_k = 0\ \forall k$ and solve it using Algorithm \ref{alg:ch_SoS_sec_TV_ADMM}.

In both \gls{sg} and \gls{tv} cases, we also define the regularization parameters $\lambda^{\text{x}}$ and $\lambda^{\text{z}}$ empirically according to the same criteria used to determine the regularization parameters in Section~\ref{sec:exp_param}. In the \gls{sg} case, we have $\lambda^{\text{x}}=12.3 \cdot 10^{12}$ and $\lambda^{\text{z}}=12.3 \cdot 10^{12}/40$, whereas in the \gls{tv} case, we have $\lambda^{\text{x}}=2.5 \cdot 10^{12}$ and $\lambda^{\text{z}}=2.5 \cdot 10^{12}/40$.

\section{Results}

\begin{figure*}[ht]
    \centering
    \centerline{\includegraphics{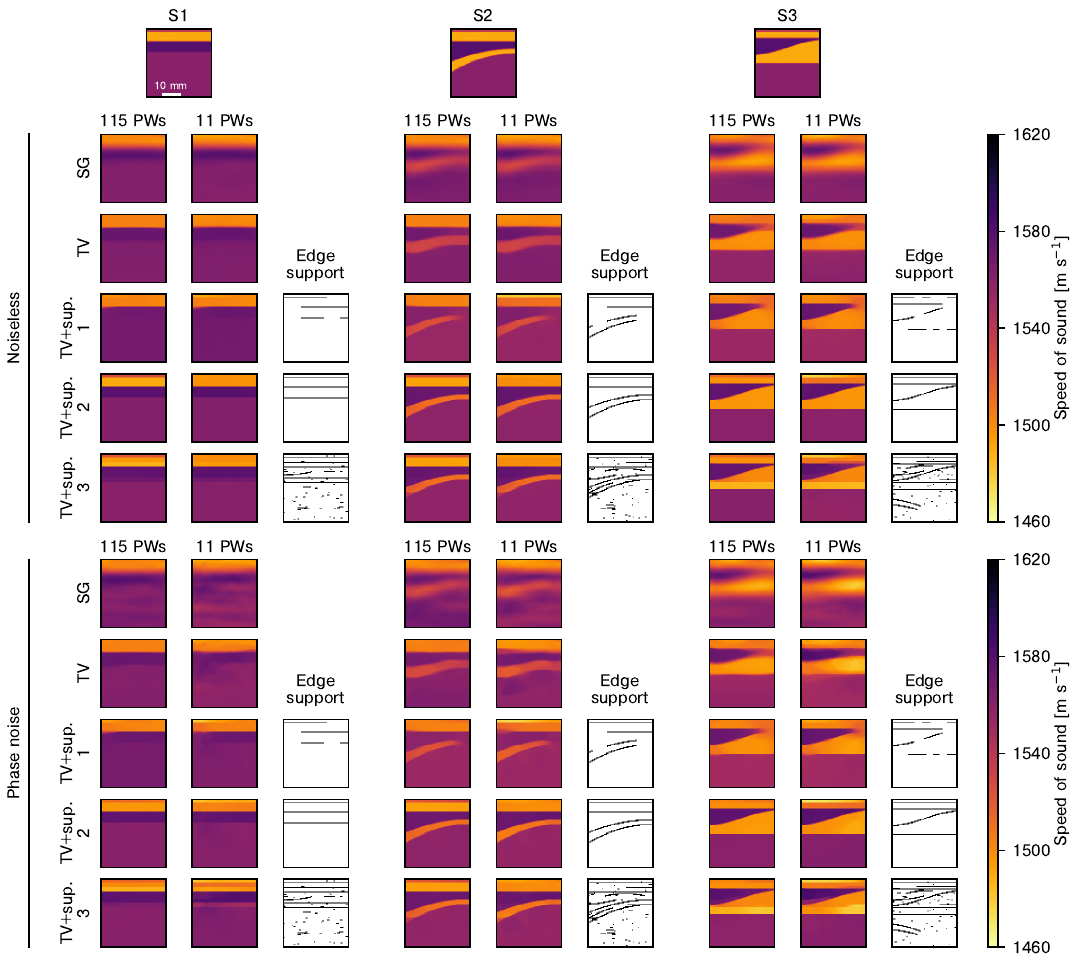}}
    \caption{Comparison between the three regularizers (\gls{sg}, \gls{tv} norm and \gls{tv} norm with partially known support) for the first three simulated datasets S1 to S3, both without and with phase noise. In each case we depict the ground-truth \gls{sos} map and the three edge support maps tested. The first edge support simulates a case where only a portion of \gls{sos} edges are detected, the second edge support represent exactly the edges of the \gls{sos} map and the third map figures a case where additional edges not corresponding to \gls{sos} transitions are detected. }
    \label{fig:fig_simulation}
\end{figure*}

\begin{figure}[ht]
    \centering
    \centerline{\includegraphics{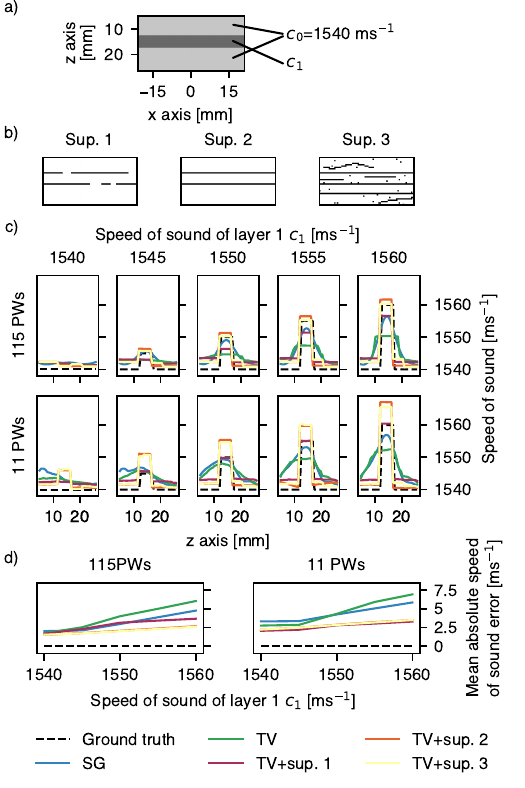}}
    \caption{Assessment of the minimum \gls{sos} contrast; a) the configuration of the reference phantom restricted to the mid field; b) the three edge support maps used in the experiments; c) profiles of ground-truth and estimated \gls{sos} evaluated at $x$=0 mm for different values of $c_1$ using the three regularizers (SG, TV norm and TV norm with partially known support) using both 115 \glspl{pw} and 11 \glspl{pw}; d) evolution of the mean absolute error with $c_1$.}
    \label{fig:fig_contrast}
\end{figure}

\begin{figure}[ht]
    \centering
    \centerline{\includegraphics{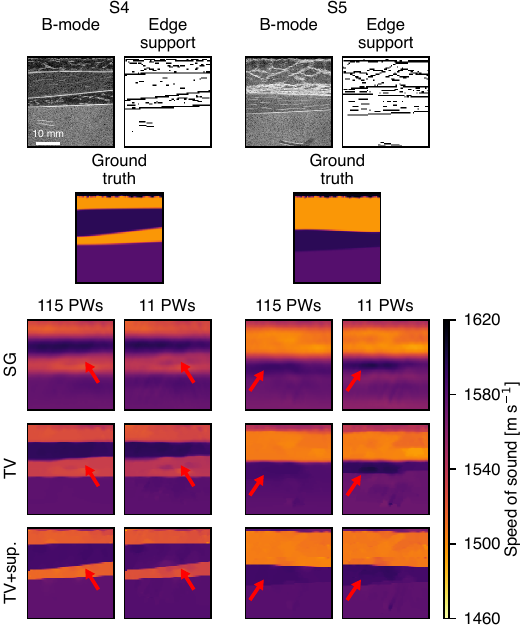}}
    \caption{Comparison between the three regularizers (\gls{sg}, \gls{tv} norm and \gls{tv} norm with partially known support) for datasets S4 and S5. For both datasets, we depict B-mode images of the medium, the corresponding edge support maps, the ground-truth \gls{sos} maps, and the estimated \gls{sos} maps using both 115 \glspl{pw} and 11 \glspl{pw}.}
    \label{fig:fig_simulation_2}
\end{figure}

\subsection{Simulated Data}

We depict in Fig.~\ref{fig:fig_simulation} a comparison of the three regularizers applied to noiseless and noisy versions of simulated datasets S1 to S3, considering the full and reduced amount of data. We display the different edge masks $\boldsymbol{q}^{\Delta s}$ used for each experiment. In addition, we present in Fig.~\ref{fig:fig_contrast} the results of the experiments designed to assess minimum contrast, whereas  Fig.~\ref{fig:fig_simulation_2} displays the results of the three regularizers applied to datasets S4 and S5 considering also both amounts of data. Finally, Table~\ref{tab:simulation} presents the median absolute error of the estimated \gls{sos} maps S1 to S5 with respect to the ground truth computed on the whole imaged medium.

The main conclusion we can draw from Fig.~\ref{fig:fig_simulation} is that the use of \gls{tv} norm with partially known support as regularizer leads, on average, to maps that are qualitatively closer to the ground truth compared to the reference techniques. This conclusion extends to quantitative measures, as can be observed in Table~\ref{tab:simulation}. In addition, we observe that \gls{tv} norm without the use of a support already allows more accurate results than the \gls{sg} approach.

\begin{table*}[ht]
	\centering
	\caption{Median absolute SoS errors (in m/s) using simulated data}
 \setlength{\tabcolsep}{4.57pt}
	\begin{tabular}{c c | c c c | c | c c c | c | c c |>{\centering\arraybackslash}p{0.0005in}}
\noalign{\hrule height 2pt}& & \multicolumn{4}{c|}{} & \multicolumn{4}{c|}{} & & &\\[-9pt]
& & \multicolumn{4}{c|}{Noiseless} & \multicolumn{4}{c|}{Phase noise} & & &\\\cline{3-10}
& & & & & & & & & & & & \\[-9pt]
& & S1 & S2 & S3 & All  & S1 & S2 & S3 & All & S4 & S5 &\\
\noalign{\hrule height 2pt} & & & & & & & & & & & & \\[-9pt] \multirow{5}{*}{\rotatebox[origin = c]{90}{115 PWs\qquad\quad\:\:\:}\:} & SG 
& \begin{tabular}{@{}c@{}}2.4\\ (1.9-10.8) \end{tabular} 
& \begin{tabular}{@{}c@{}}8.6\\ (4.7-24.8) \end{tabular} 
& \begin{tabular}{@{}c@{}}9.3\\ (0.9-23.3) \end{tabular} 
& \begin{tabular}{@{}c@{}}6.1\\ (2.1-20.7) \end{tabular} 
& \begin{tabular}{@{}c@{}}4.8\\ (2.3-10.8) \end{tabular} 
& \begin{tabular}{@{}c@{}}10.2\\ (6.8-25.0) \end{tabular} 
& \begin{tabular}{@{}c@{}}7.4\\ (1.8-22.8) \end{tabular} 
& \begin{tabular}{@{}c@{}}7.6\\ (3.0-20.4) \end{tabular}
& \begin{tabular}{@{}c@{}}19.3\\ (5.0-32.2) \end{tabular}
& \begin{tabular}{@{}c@{}}15.2\\ (7.1-16.2) \end{tabular} &
\\\cline{2-13} & & & & & & & & & & & & \\[-9pt]& TV 
& \begin{tabular}{@{}c@{}}3.5\\ (2.7-7.4) \end{tabular} 
& \begin{tabular}{@{}c@{}}6.1\\ (1.8-18.3) \end{tabular} 
& \begin{tabular}{@{}c@{}}8.4\\ (3.3-12.6) \end{tabular} 
& \begin{tabular}{@{}c@{}}3.7\\ (2.7-14.4) \end{tabular} 
& \begin{tabular}{@{}c@{}}2.1\\ (1.7-8.0) \end{tabular} 
& \begin{tabular}{@{}c@{}}3.7\\ (2.0-16.6) \end{tabular} 
& \begin{tabular}{@{}c@{}}7.3\\ (4.5-12.2) \end{tabular} 
& \begin{tabular}{@{}c@{}}5.5\\ (2.1-14.6) \end{tabular}
& \begin{tabular}{@{}c@{}}17.0\\ (2.6-31.0) \end{tabular}
& \begin{tabular}{@{}c@{}}10.5\\ (2.2-11.6) \end{tabular} &
\\\cline{2-13} & & & & & & & & & & & & \\[-9pt]&
TV+sup. 1
& \begin{tabular}{@{}c@{}}7.4\\ (7.3-10.8) \end{tabular} 
& \begin{tabular}{@{}c@{}}6.1\\ (6.0-19.0) \end{tabular} 
& \begin{tabular}{@{}c@{}}7.9\\ (7.3-12.5) \end{tabular} 
& \begin{tabular}{@{}c@{}}7.4\\ (7.3-14.0) \end{tabular} 
& \begin{tabular}{@{}c@{}}6.7\\ (6.2-8.9) \end{tabular} 
& \begin{tabular}{@{}c@{}}7.5\\ (6.8-21.7) \end{tabular} 
& \begin{tabular}{@{}c@{}}10.2\\ (10.1-12.4) \end{tabular} 
& \begin{tabular}{@{}c@{}}9.2\\ (6.8-13.9) \end{tabular} & & &
\\\cline{2-10} & & & & & & & & & & & & \\[-9pt]&
TV+sup. 2 
& \begin{tabular}{@{}c@{}}1.3\\ (1.3-1.4) \end{tabular} 
& \begin{tabular}{@{}c@{}} \textbf{2.1}\\ (2.1-7.6) \end{tabular} 
& \begin{tabular}{@{}c@{}}4.0\\ (0.2-6.8) \end{tabular} 
& \begin{tabular}{@{}c@{}}\textbf{2.1}\\ (1.3-6.6) \end{tabular} 
& \begin{tabular}{@{}c@{}}1.0\\ (0.9-4.1) \end{tabular} 
& \begin{tabular}{@{}c@{}}\textbf{3.0}\\ (2.3-9.7) \end{tabular} 
& \begin{tabular}{@{}c@{}}2.9\\ (0.2-6.9) \end{tabular} 
& \begin{tabular}{@{}c@{}}\textbf{2.3}\\ (0.9-6.7) \end{tabular}
& \begin{tabular}{@{}c@{}}\textbf{12.8}\\ (3.6-17.5) \end{tabular}
& \begin{tabular}{@{}c@{}}\textbf{7.4}\\ (0.5-10.3) \end{tabular} &
\\\cline{2-10} & & & & & & & & & & & & \\[-9pt]&
TV+sup. 3 
& \begin{tabular}{@{}c@{}}\textbf{0.4}\\ (0.3-4.3) \end{tabular} 
& \begin{tabular}{@{}c@{}} 3.0\\ (2.9-8.7) \end{tabular} 
& \begin{tabular}{@{}c@{}}\textbf{2.5}\\ (1.0-8.6) \end{tabular} 
& \begin{tabular}{@{}c@{}}2.9\\ (1.0-6.6) \end{tabular} 
& \begin{tabular}{@{}c@{}}\textbf{0.8}\\ (0.7-3.1) \end{tabular} 
& \begin{tabular}{@{}c@{}}3.6\\ (3.2-9.6) \end{tabular} 
& \begin{tabular}{@{}c@{}}\textbf{1.9}\\ (1.4-10.1) \end{tabular} 
& \begin{tabular}{@{}c@{}}3.1\\ (1.4-8.7) \end{tabular} & & &
\\\cline{1-13} & & & & & & & & & & & & \\[-9pt]\multirow{5}{*}{\rotatebox[origin = c]{90}{11 PWs\qquad\qquad}\:} & SG 
& \begin{tabular}{@{}c@{}}2.6\\ (1.4-10.0) \end{tabular} 
& \begin{tabular}{@{}c@{}}7.4\\ (4.7-24.6) \end{tabular} 
& \begin{tabular}{@{}c@{}}8.8\\ (1.0-23.3) \end{tabular} 
& \begin{tabular}{@{}c@{}}6.0\\ (2.0-20.5) \end{tabular} 
& \begin{tabular}{@{}c@{}}5.7\\ (2.7-11.8) \end{tabular} 
& \begin{tabular}{@{}c@{}}9.2\\ (4.4-26.0) \end{tabular} 
& \begin{tabular}{@{}c@{}}8.8\\ (4.3-22.9) \end{tabular} 
& \begin{tabular}{@{}c@{}}7.5\\ (3.8-21.0) \end{tabular}
& \begin{tabular}{@{}c@{}}20.2\\ (4.3-34.3) \end{tabular}
& \begin{tabular}{@{}c@{}}17.5\\ (8.6-21.3) \end{tabular} &
\\\cline{2-13} & & & & & & & & & & & & \\[-9pt]& TV 
& \begin{tabular}{@{}c@{}}4.4\\ (3.2-7.4) \end{tabular} 
& \begin{tabular}{@{}c@{}}6.6\\ (2.8-16.7) \end{tabular} 
& \begin{tabular}{@{}c@{}}8.6\\ (2.1-12.7) \end{tabular} 
& \begin{tabular}{@{}c@{}}5.2\\ (3.0-12.1) \end{tabular} 
& \begin{tabular}{@{}c@{}}2.8\\ (2.0-9.7) \end{tabular} 
& \begin{tabular}{@{}c@{}}6.1\\ (2.9-22.5) \end{tabular} 
& \begin{tabular}{@{}c@{}}9.4\\ (6.9-16.4) \end{tabular} 
& \begin{tabular}{@{}c@{}}7.9\\ (2.7-14.4) \end{tabular}
& \begin{tabular}{@{}c@{}}18.3\\ (1.8-33.8) \end{tabular}
& \begin{tabular}{@{}c@{}}13.4\\ (6.6-15.2) \end{tabular} &
\\\cline{2-13} & & & & & & & & & & & & \\[-9pt]& \:
TV+sup. 1
& \begin{tabular}{@{}c@{}}8.2\\ (8.0-9.0) \end{tabular} 
& \begin{tabular}{@{}c@{}}5.9\\ (5.0-20.7) \end{tabular} 
& \begin{tabular}{@{}c@{}}8.9\\ (8.4-11.4) \end{tabular} 
& \begin{tabular}{@{}c@{}}8.5\\ (7.9-13.7) \end{tabular} 
& \begin{tabular}{@{}c@{}}3.6\\ (3.0-9.3) \end{tabular} 
& \begin{tabular}{@{}c@{}}7.9\\ (7.1-21.3) \end{tabular} 
& \begin{tabular}{@{}c@{}}9.0\\ (7.8-12.2) \end{tabular} 
& \begin{tabular}{@{}c@{}}8.2\\ (5.8-15.0) \end{tabular} & & &
\\\cline{2-10} & & & & & & & & & & & & \\[-9pt]&
TV+sup. 2 
& \begin{tabular}{@{}c@{}}1.9\\ (1.7-2.3) \end{tabular} 
& \begin{tabular}{@{}c@{}}\textbf{2.8}\\ (2.5-10.4) \end{tabular} 
& \begin{tabular}{@{}c@{}}\textbf{2.9}\\ (0.8-7.8) \end{tabular} 
& \begin{tabular}{@{}c@{}}\textbf{2.5}\\ (1.7-7.5) \end{tabular} 
& \begin{tabular}{@{}c@{}}\textbf{1.4}\\ (0.8-5.2) \end{tabular} 
& \begin{tabular}{@{}c@{}}\textbf{5.0}\\ (4.8-9.9) \end{tabular} 
& \begin{tabular}{@{}c@{}}\textbf{1.9}\\ (1.4-7.8) \end{tabular} 
& \begin{tabular}{@{}c@{}}\textbf{4.3}\\ (1.4-9.1) \end{tabular}
& \begin{tabular}{@{}c@{}}\textbf{14.0}\\ (0.6-24.9) \end{tabular}
& \begin{tabular}{@{}c@{}}\textbf{9.1}\\ (3.0-12.0) \end{tabular} &
\\\cline{2-10} & & & & & & & & & & & & \\[-9pt]& 
TV+sup. 3 
& \begin{tabular}{@{}c@{}}\textbf{1.1}\\ (0.9-2.6) \end{tabular} 
& \begin{tabular}{@{}c@{}}3.1\\ (2.8-10.4) \end{tabular} 
& \begin{tabular}{@{}c@{}}4.4\\ (3.2-12.1) \end{tabular} 
& \begin{tabular}{@{}c@{}}3.2\\ (2.3-9.2) \end{tabular} 
& \begin{tabular}{@{}c@{}}2.6\\ (2.3-5.7) \end{tabular} 
& \begin{tabular}{@{}c@{}}\textbf{5.0}\\ (4.8-10.7) \end{tabular} 
& \begin{tabular}{@{}c@{}}2.1\\ (0.6-10.4) \end{tabular} 
& \begin{tabular}{@{}c@{}}4.7\\ (2.3-9.8) \end{tabular}  & & &
\\\noalign{\hrule height 2pt} \multicolumn{13}{l}{} \\[-9pt]
	\end{tabular}
	\label{tab:simulation}
\end{table*}

However, major differences can be observed when different support maps are employed. First of all, the proposed approach leads in practice to biases in the \gls{sos} map when the reduced edge support map is used. Such biases can be observed visually and quantitatively in Table~\ref{tab:simulation}, where we notice that the median error can even be worse than TV norm without support.

In contrast, the proposed regularizer leads to \gls{sos} estimates that correspond to a high degree to the ground-truth \gls{sos} maps when the second and third support maps are used. A qualitative improvement can first be highlighted. The latter is especially noticeable in case S2, where the structure of the intermediate low-\gls{sos} layer is completely different with or without partially known support. There is also a quantitative improvement since the error achieved by the proposed regularizer is consistently lower than the common \gls{tv} regularizer.
Moreover, we must highlight that only limited qualitative and quantitative differences can be observed between results generated with the second and third support maps. In particular, it demonstrates the stability of the proposed regularizer to noise in the support map $\boldsymbol{q}^{\Delta s}$. 

As it can be seen in Fig.~\ref{fig:fig_contrast}, the observations highlighted above can be mostly extended to the results of our assessment of minimum contrast. The location of layer 1 is better recovered with the proposed regularizer, irrespective of the support map considered. When 115 \glspl{pw} are used,
the proposed regularizer with supports 2 and 3 consistently outperform the other SG and TV regularizers in terms of mean absolute error, as well as the proposed method using support 1, even though a \SI{2}{\meter \per \second} bias exists irrespective of the method used. With respect to the 11-\gls{pw} experiments, the \gls{sos} profiles all suffer from a larger amount of noise. The \gls{sos} bias within layer 1 increases with respect to the 115-\gls{pw} case when the proposed regularizer uses supports 2 and 3, such that the errors achieved by the three support maps are similar across the different values of $c_1$. Nonetheless, the proposed regularizer still largely outperforms the two reference approaches.

Regarding datasets S4 and S5, the proposed prior consistently outperforms \gls{sg} and \gls{tv} norm for both amounts of data. This can be seen qualitatively in Fig.~\ref{fig:fig_simulation_2}, where the topologies of the estimated \gls{sos} maps are closer to the ones of the ground-truth maps, or quantitatively in the right quarter of Table~\ref{tab:simulation}, where the median absolute error is consistently lower with the proposed regularizer. Due to the more complex \gls{sos} and echogenicity distributions, the median absolute errors are overall higher for cases S4 and S5 than for cases S1 to S3, and the relative improvement provided by the proposed regularizer is lower on average. Nevertheless,  experiments S4 and S5 show that the negative effect of the spatial shift in the edge support map remains limited and do not threaten the applicability of \gls{tv} norm with partially-known support. Moreover, the positive effects of the proposed regularizer can be seen in its capacity to improve the reconstruction of the fat layers in case S4 (see Fig.~\ref{fig:fig_simulation_2}) or to distinguish the median muscular layer in case S5 even with a \gls{sos} lower than the ground-truth---where the layer appears thinner or disappears completely with the \gls{sg} or \gls{tv} regularizers. In particular, these last two observations show that the proposed regularizer is able to improve \gls{sos} contrast compared to previous state-of-the-art regularizers, while being able to distinguish layers with a lower \gls{sos} difference.

\subsection{In Vivo Data}

\begin{figure*}[!t]
    \centering
    \centerline{\includegraphics{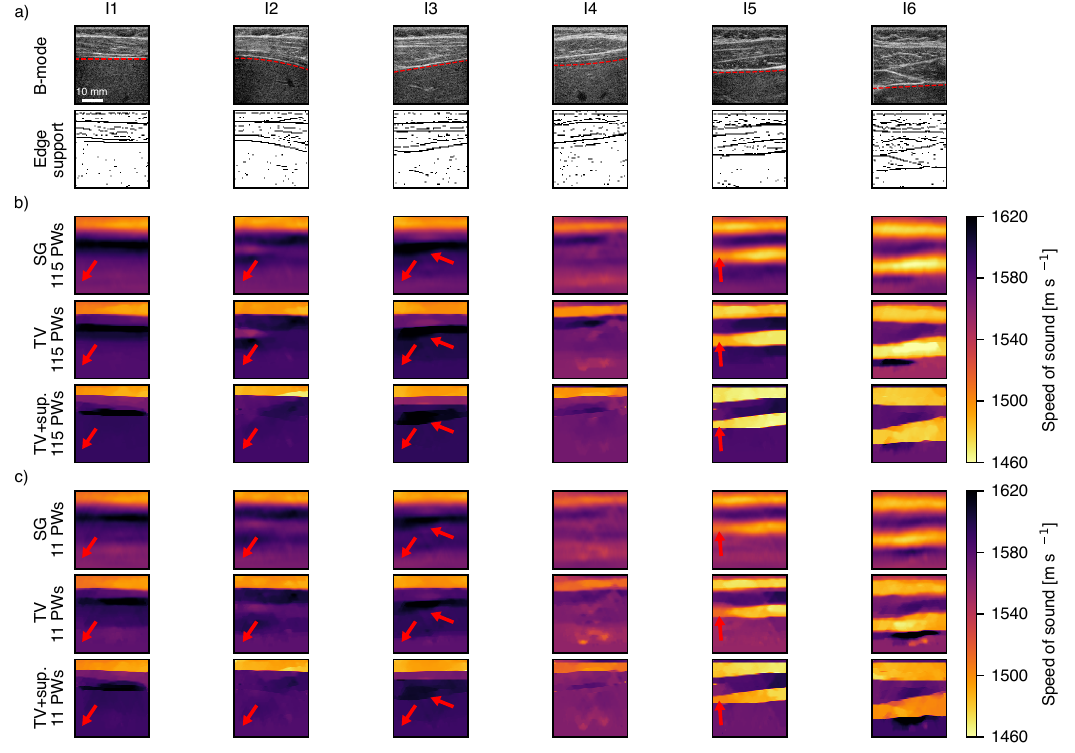}}
    \caption{In vivo comparison of the three regularizers; a) B-mode images of the medium with the estimated mask indicating the support of the edges in the \gls{sos} map; b) \gls{sos} maps reconstructed from 115-\gls{pw} data, using a SG regularizer, a TV norm regularizer and the proposed regularizer; c) \gls{sos} maps reconstructed from 11-\gls{pw} data.}
    \label{fig:invivo}
\end{figure*}

\begin{table}
	\centering
	\caption{In vivo median SoS (in m/s) in the liver}
    \setlength{\tabcolsep}{3.55pt}
	\begin{tabular}{ c  c |c c  c  c  c c | >{\centering\arraybackslash}p{0.0001in}}
						\noalign{\hrule height 2pt}& & & & & & & & \\[-9pt] & & I1 & I2 & I3 & I4 & I5 & I6 & \\ \noalign{\hrule height 2pt} & & & & & & & & \\[-9pt]\multirow{3}{*}{\rotatebox[origin = c]{90}{115 PWs\quad}\:} & SG 
					& \begin{tabular}{@{}c@{}}1571.4\\ (12.8) \end{tabular} 
					& \begin{tabular}{@{}c@{}}1573.2\\ (14.6) \end{tabular} 
					& \begin{tabular}{@{}c@{}}1581.3\\ (15.6) \end{tabular} 
					& \begin{tabular}{@{}c@{}}1561.0\\ (16.9) \end{tabular} 
					& \begin{tabular}{@{}c@{}}1573.5\\ (12.6) \end{tabular} 
					& \begin{tabular}{@{}c@{}}1565.4\\ (10.6) \end{tabular} 
					& \\\cline{2-9} & & & & & & & & \\[-9pt]& TV 
					& \begin{tabular}{@{}c@{}}1583.0\\ (5.5) \end{tabular} 
					& \begin{tabular}{@{}c@{}}1587.3\\ (5.7) \end{tabular} 
					& \begin{tabular}{@{}c@{}}1595.6\\ (6.5) \end{tabular} 
					& \begin{tabular}{@{}c@{}}1561.9\\ (10.6) \end{tabular} 
					& \begin{tabular}{@{}c@{}}1586.3\\ (7.2) \end{tabular} 
					& \begin{tabular}{@{}c@{}}1551.3\\ (0.8) \end{tabular} 
					& \\\cline{2-9} & & & & & & & & \\[-9pt]&
                    TV+sup.
					& \begin{tabular}{@{}c@{}}1593.1\\ (\textbf{0.8}) \end{tabular} 
					& \begin{tabular}{@{}c@{}}1587.9\\ (\textbf{1.5}) \end{tabular} 
					& \begin{tabular}{@{}c@{}}1594.8\\ (\textbf{1.1}) \end{tabular} 
					& \begin{tabular}{@{}c@{}}1569.0\\ (\textbf{2.8}) \end{tabular} 
					& \begin{tabular}{@{}c@{}}1583.7\\ (\textbf{0.7}) \end{tabular} 
					& \begin{tabular}{@{}c@{}}1573.6\\ (\textbf{0.1}) \end{tabular} 
					& \\\hline & & & & & & & & \\[-9pt]\multirow{3}{*}{\rotatebox[origin = c]{90}{11 PWs\quad\:\,}\:} & SG 
					& \begin{tabular}{@{}c@{}}1568.7\\ (22.3) \end{tabular} 
					& \begin{tabular}{@{}c@{}}1575.6\\ (17.9) \end{tabular} 
					& \begin{tabular}{@{}c@{}}1584.2\\ (22.0) \end{tabular} 
					& \begin{tabular}{@{}c@{}}1560.0\\ (15.9) \end{tabular} 
					& \begin{tabular}{@{}c@{}}1558.1\\ (6.6) \end{tabular} 
					& \begin{tabular}{@{}c@{}}1561.5\\ (16.8) \end{tabular} 
					& \\\cline{2-9} & & & & & & & & \\[-9pt]& TV 
					& \begin{tabular}{@{}c@{}}1578.3\\ (15.6) \end{tabular} 
					& \begin{tabular}{@{}c@{}}1581.9\\ (12.1) \end{tabular} 
					& \begin{tabular}{@{}c@{}}1592.3\\ (17.1) \end{tabular} 
					& \begin{tabular}{@{}c@{}}1557.5\\ (10.7) \end{tabular} 
					& \begin{tabular}{@{}c@{}}1555.6\\ (3.6) \end{tabular} 
					& \begin{tabular}{@{}c@{}}1560.3\\ (4.9) \end{tabular} 
					& \\\cline{2-9} & & & & & & & & \\[-9pt]&
                    TV+sup.
					& \begin{tabular}{@{}c@{}}1589.8\\ (\textbf{3.7}) \end{tabular} 
					& \begin{tabular}{@{}c@{}}1583.3\\ (\textbf{3.0}) \end{tabular} 
					& \begin{tabular}{@{}c@{}}1591.4\\ (\textbf{4.1}) \end{tabular} 
					& \begin{tabular}{@{}c@{}}1562.8\\ (\textbf{2.4}) \end{tabular} 
					& \begin{tabular}{@{}c@{}}1571.4\\ (\textbf{0.9}) \end{tabular} 
					& \begin{tabular}{@{}c@{}}1589.5\\ (\textbf{1.6}) \end{tabular} 
					& \\\noalign{\hrule height 2pt} 
	\end{tabular}
	\label{tab:invivo}
\end{table}

We depict in Fig.~\ref{fig:invivo} the results of the proposed method applied to datasets I1 to I6. We use both 115 (Fig. ~\ref{fig:invivo}b) and 11  (Fig. ~\ref{fig:invivo}c) \gls{pw} insonifications and present the results of the three regularizers. In Fig.~\ref{fig:invivo}a, we display B-mode images of the medium, as well as the edge support maps $\boldsymbol{q}^{\Delta s}$ computed with the method outlined in Section~\ref{sec:meth_ridge}. In addition, we present in Table~\ref{tab:invivo} the median \gls{sos} in the liver estimated in each case with each regularizer, with the interquartile distance provided between parentheses. The top boundary of the liver is indicated by a red dashed line in Fig.~\ref{fig:invivo}a.

Regarding first cases I1 to I3, we observe in Fig.~\ref{fig:invivo}c that areas with lower \gls{sos} appear in the bottom half of the liver for both SG and TV regularizers when we use 11-\gls{pw} data. In contrast, they do not appear in the 115-\gls{pw} results, and no corresponding structure can be observed in the B-mode images. The low-\gls{sos} areas, however, are not present anymore in the deep liver with the proposed regularizer. More generally, the \gls{sos} variability in the liver---as measured by the interquartile distance in Table~\ref{tab:invivo}---is significantly reduced by the proposed regularizer compared to the two references ones. 

Focusing now on I5, we can see that the \gls{sos} in the liver using 11 \glspl{pw} is inferior to the 115-\gls{pw} results with the reference regularizers. In contrast, the difference between the two amounts of \glspl{pw} is reduced by the proposed method. Nonetheless, we observe a loss of \gls{sos} contrast when the amount of data is reduced irrespective of the regularizer, as it can be seen in both cases I5 and I6.

Qualitatively, we see that the transitions between the different layers of the medium are more well-defined with the proposed regularizer compared to the TV norm. The improvement is especially visible with regard to the upper boundary of the liver in I3, or in the left part of the middle layer of I5 with 11 \glspl{pw}. However, some of the superficial layers of I2 and I4---even though they are clearly visible in the B-mode image---appear to share approximately the same \gls{sos} as the liver. This may be a sign that the proposed method fails to recover some of the details of the physical \gls{sos} distribution.

Finally, we can highlight that the liver \gls{sos} achieved by the proposed regularizer is, on average, closer to the average \gls{sos} of a healthy human liver (i.e., between \SI{1580}{\meter \per \second} and \SI{1600}{\meter \per \second} \cite{mast_empirical_2000}) than with the reference regularizers. Indeed, the average \gls{sos} achieved is \SI{1583.7}{\meter \per \second} for 115 \glspl{pw} and \SI{1581.4}{\meter \per \second} for 11 \glspl{pw}, whereas they equal \SI{1571}{\meter \per \second} and \SI{1568}{\meter \per \second} in the SG case, and \SI{1577.6}{\meter \per \second} and \SI{1571}{\meter \per \second} in the TV case. Moreover, the \gls{sos} values estimated in the other tissues (fat and muscle) are also within their expected range for a healthy individual. Overall, in vivo results tend to show that the edge support detected by the ridge detection algorithm corresponds to anatomical boundaries between tissues. However, these observations must be put into perspective considering the absence of ground truth.

\section{Discussion}

The benefits of the proposed regularizer on the estimation of liver \gls{sos} are most apparent when using a reduced number of insonifications. In such cases, the proposed regularizer alleviates \gls{sos} variations within the liver and recovers quantitative in vivo liver \gls{sos} values that more closely match the average values observed in healthy individuals. This demonstrates the potential of the regularizer to enhance the accuracy of pulse-echo \gls{sos} imaging when using ultra-portable transducers. These devices are often constrained by limited electrical power and restricted data storage or transmission capacity, making it essential to minimize the number of required insonifications.

Regardless of the number of insonifications used, another effect of the regularizer is to better estimate the details of the \gls{sos} map. This effect is especially noticeable on simulation results---for example with S2 in Fig.~\ref{fig:fig_simulation}. As expected, the in vivo results fit more closely the tissue structure observed in the B-mode image.

However, the proposed regularizer is effective only if a very large proportion of the \gls{sos} edges is detected in the B-mode image. If it is not the case, significant errors can occur in the estimated \gls{sos} distribution. Such phenomena can be observed in the simulated results using reduced edge support map (see Fig.~\ref{fig:fig_simulation}, TV+sup. 1). In addition, the top boundary of the liver is not completely detected in example I4 (see Fig.~\ref{fig:invivo}a), and the estimated local \gls{sos} is almost uniform up to the interface located above the boundary of the liver. Thus, the proposed regularizer alleviates a potential advantage of \gls{sos} imaging, namely to distinguish structures not visible in B-mode images.

On the other hand, the proposed prior enforces more strongly tissue-wise uniformity compared to the SG or TV regularizers. Therefore, the risk of artifacts in the recovered \gls{sos} maps is strongly alleviated---as can be seen in the bottom layer of I6, for example. Factoring in the reduction in cross-tissue bias, we argue that the gain in robustness allowed by the proposed regularizer outweigh the risk of not detecting a structure indistinguishable in the B-mode image, especially in liver imaging. Furthermore, the estimation of the edge support map can be improved in the future, both specifically for liver imaging and to extend the method to other applications. For example, we could factor in variations in mean echogenicity as indications of possible \gls{sos} edges. This could be especially useful for the application of our method to breast imaging, where the boundaries of breast tumors typically do not appear as bright interfaces in B-mode images \cite{ruby_breast_2019}.

One notable drawback of TV-norm priors---including the proposed TV norm with partially known support---with respect to quadratic regularizers like SG is the increase in computation time. Indeed, Algorithm~\ref{alg:ch_SoS_sec_TV_ADMM} requires one matrix-vector product with $[\boldsymbol{A}^T\boldsymbol{A}]^{-1}$ per iteration, whereas the resolution of a quadratic problem requires the equivalent of a single iteration. With our current CPU implementation, each iteration of Algorithm~\ref{alg:ch_SoS_sec_TV_ADMM} requires around \SI{750}{\milli \second} on an Intel Xeon E5-2630 Processor (Intel, Santa Clara, CA, USA). Moreover, Algorithm~\ref{alg:ch_SoS_sec_TV_ADMM} typically takes between 20 and 120 iterations to converge with large variations between experiments, irrespective of if a support map is used, namely its computation time is on the order of a minute. Nonetheless, we believe that a GPU implementation of Algorithm~\ref{alg:ch_SoS_sec_TV_ADMM} would significantly improve the time per iteration, whereas the parameters of Algorithm~\ref{alg:ch_SoS_sec_TV_ADMM} can be adapted to reduce the number of iterations to make it compatible with real-time imaging. In addition, our current CPU implementation of the ridge detection method takes \SI{2.3}{\second}. Since the ridge detection method is fully parallelizable, a GPU implementation would also significantly reduce computation time.

We use herein the proposed regularizer in conjunction with the aberration phase-shift estimation method introduced in \cite{beuret_windowed_2023-1} designed for linear transducers. However, the regularizer can be readily implemented with other aberration phase shift (respectively, aberration delay) estimation methods or other types of ultrasound transducers. Indeed, a large set of pulse-echo \gls{sos} imaging methods (e.g., \cite{sanabria_speed--sound_2018, ali_sound_2023, stahli_improved_2020, jaeger2022pulse}) comprises the resolution of an inverse problem involving a convex data term and a linear measurement model adapted to the considered imaging configuration. The measurement model $\boldsymbol{M}$ used herein can thus be straightforwardly replaced by the model of interest. The proximal operator in Algorithm \ref{alg:ch_SoS_sec_TV_ADMM} can also be adapted to other distances than the squared $\ell_2$ norm. Finally, the regularization parameters and the ridge detection parameters must be adapted to the new imaging configuration. In particular, convex transducers are the favored tools in practice for liver imaging and the regularizer can be readily used with current methods relying on convex probes (e.g., \cite{jaeger2022pulse}).
With further adaptations, the regularizer can also be used with other approaches introduced recently such as differentiable beamforming \cite{simson2025ultrasound}. Even though our experiments tend to show that the errors caused by shifts in the apparent displacement of the ridges in B-mode images are limited, applying the proposed regularizer with differentiable beamforming and dynamically updating the edge support map would cancel such errors entirely.

Finally, an important question about the proposed technique concerns its in vivo repeatability. It is currently challenging to assess if the differences observed between the in vivo estimates of liver \gls{sos} are caused by inhomogeneities in the liver, expected statistical noise, or a systematic bias that would be prejudicial to diagnosis. A future pre-clinical study considering data from several patients---including patients suffering from \gls{masld}---is necessary to properly assess the repeatability of the technique in vivo and quantify more precisely its usefulness in a clinical context.

\section{Conclusion}

We propose in this article a novel regularizer for speed-of-sound imaging able to automatically factor in information from B-mode images of the insonified medium. Our results support an improvement in robustness of distributed pulse-echo \gls{sos} estimation in liver imaging compared to state-of-the-art priors, especially when the amount of data is small. In particular, this improvement is promising for the deployment of distributed pulse-echo \gls{sos} imaging methods to ultra-portable transducers since they are likely to be limited in their operating power or data transmission capabilities. More generally, we believe that the proposed principle of TV norm with partially known support allows for a versatile \gls{sos} imaging regularization strategy that can be adapted to different use cases and clinical applications.

\appendix
\section*{Appendix}

\label{sec:ch_prox_sec_lp}

According to the separability of proximal operators \cite{combettes2011proximal}, the computation of $\Prox_{\mathcal{F}}$ can be split into a series of sub-problems. For the first term of function $\mathcal{F}$, we have
\begin{equation}
\label{eq:prox_app_1}
    \left[\Prox_{\mathcal{F}} \big( \boldsymbol{\zeta} \big)\right]_k =
        \underset{\hat{\zeta}_1}{\argmin} \frac{1}{2} \big[ \hat{\zeta}_1\big]^2 + \frac{1}{2} \big[ \hat{\zeta}_1 - \big[\zeta_1\big]_k\big]^2,
\end{equation}
for $k=1,...,N^{\text{d}}$. Problem \eqref{eq:prox_app_1} admits a straightforward closed-form solution
\begin{equation}
\label{eq:prox_app_2}
    \left[\Prox_{\mathcal{F}} \big( \boldsymbol{\zeta} \big)\right]_k =
        \frac{1}{2} \big[\zeta_1\big]_k.
\end{equation}
For the second term of $\mathcal{F}$, the proximal operator amounts to a series of sub-problems
\begin{multline}
\label{eq:prox_app_3}
    \begin{bmatrix}
         \left[\Prox_{\mathcal{F}} \big( \boldsymbol{\zeta} \big)\right]_{N^{\text{d}} + k}\\
         \left[\Prox_{\mathcal{F}} \big( \boldsymbol{\zeta} \big)\right]_{N^{\text{d}} + N^{\text{r}} + k}
    \end{bmatrix} =
        \underset{\hat{\zeta}_2, \hat{\zeta}_3}{\argmin} \frac{1}{2} \big[ \hat{\zeta}_2 - \big[\zeta_2\big]_k \big]^2 +\\ \frac{1}{2} \big[ \hat{\zeta}_3 - \big[\zeta_3\big]_k\big]^2 + \big[1 - q^{\Delta s}_k \big] \sqrt{ \big[\hat{\zeta}_2\big]^2 + \big[\hat{\zeta}_3\big]^2},
\end{multline}
with $k=1,...,N^{\text{r}}$. If we rewrite this subproblem in polar coordinates, it follows that problem \eqref{eq:prox_app_3} can be expressed as
\begin{equation}
    \begin{bmatrix}
         \left[\Prox_{\mathcal{F}} \big( \boldsymbol{\zeta} \big)\right]_{N^{\text{d}} + k}\\
         \left[\Prox_{\mathcal{F}} \big( \boldsymbol{\zeta} \big)\right]_{N^{\text{d}} + N^{\text{r}}+ k}
    \end{bmatrix} = \begin{cases}
        \frac{\rho^{\prime}}{\rho_k} \begin{bmatrix}
            \big[ \zeta_2]_k\\
            \big[ \zeta_3]_k 
        \end{bmatrix} &\text{ if } q^{\Delta s} = 0\\
        \begin{bmatrix}
            \big[ \zeta_2]_k\\
            \big[ \zeta_3]_k 
        \end{bmatrix}  &\text{ if } q^{\Delta s} = 1,
    \end{cases}
\end{equation}
with
\begin{equation}
    \rho_k = \sqrt{\big[ \zeta_2]_k^2 + \big[ \zeta_3]_k^2},
\end{equation}
and 
\begin{equation}
    \rho^{\prime} = \underset{\hat{\rho}\geq 0}{\argmin} \frac{1}{2}\big[ \hat{\rho} - \rho_k \big]^2 + \hat{\rho}.
\end{equation}
This last sub-problem admits a simple closed-form solution expressed as
\begin{equation}
    \rho^{\prime} = \max \left\{ \rho_k - 1 , 0 \right\}.
\end{equation}

\section*{Acknowledgment}

The authors would like to thank Baptiste Hériard-Dubreuil for the discussions regarding the article and Adrien Besson for his proofreading.

\bibliographystyle{ACM-Reference-Format}
\bibliography{IEEEabrv,references}

\end{document}